\documentclass[letterpaper, 10 pt, conference]{ieeeconf}  

\IEEEoverridecommandlockouts                              

\usepackage{xcolor}
\usepackage{color,colortbl}
\definecolor{Gray}{gray}{0.9} 
\usepackage{lipsum}
\usepackage{mwe} 

\title{\LARGE \bf
CNN-LSTM Based Multimodal MRI and Clinical Data Fusion for\\ Predicting Functional Outcome in Stroke Patients
}

\author{Nima Hatami$^{1}$, Tae-Hee Cho$^{2,3}$, Laura Mechtouff$^{2,3}$, Omer Faruk Eker$^{1,4}$, David Rousseau$^{5}$ and Carole Frindel$^{1}$
\thanks{$^{1}$CREATIS,
Université Lyon1, CNRS UMR5220, INSERM U1206, INSA-Lyon, 69621 Villeurbanne, France {\tt\small carole.frindel@creatis.insa-lyon.fr}}%
\thanks{$^{2}$Department of Vascular Neurology, Hospices Civils de Lyon, France}%
\thanks{$^{3}$CarMeN, INSERM U1060, INRA U1397, Université Lyon 1, INSA de Lyon}%
\thanks{$^{4}$Department of Interventional Neuroradiology, Hospices Civils de Lyon, France}%
\thanks{$^{5}$LARIS, UMR IRHS INRA, Universite d’Angers, France}
}

\begin{document}

\maketitle
\thispagestyle{empty}
\pagestyle{empty}

\begin{abstract}
Clinical outcome prediction plays an important role in stroke patient management. 
From a machine learning point-of-view, one of the main challenges is dealing with heterogeneous data at patient admission, i.e. the image data which are multidimensional and the clinical data which are scalars. In this paper, a multimodal convolutional neural network - long short-term memory (CNN-LSTM) based ensemble model is proposed. For each MR image module, a dedicated network provides preliminary prediction of the clinical outcome using the modified Rankin scale (mRS). 
The final mRS score is obtained by merging the preliminary probabilities of each module dedicated to a specific type of MR image weighted by the clinical metadata, here age or the National Institutes of Health Stroke Scale (NIHSS). The experimental results demonstrate that the proposed model surpasses the baselines and offers an original way to automatically encode the spatio-temporal context of MR images in a deep learning architecture. The highest AUC ($0.77$) was achieved for the proposed model with NIHSS. 
\newline

\indent \textit{Clinical relevance} — 
We present the first deep learning approach predicting the clinical outcome of stroke patients treated by mechanical thrombectomy which integrates imaging data at the voxel level with key clinical metadata. Combining clinical and imaging data to evaluate the potential benefit from therapy closely mirrors the clinical decision process. Our promising results suggest our predictive model could assist in acute stroke management.

\end{abstract}

\section{INTRODUCTION}

Ischemic stroke is a leading cause of acquired disability, dementia and mortality worldwide \cite{gorelick2019global}. Several machine learning-based studies have been devoted to the prediction of the final stroke lesion \cite{winzeck2018isles,nielsen2018prediction,yu2020use,Debs21}, but still very few to the question of clinical outcome \cite{van2018predicting,ramos2020predicting}. The clinical outcome is usually measured by the modified Rankin scale (mRS) which grades the degree of disability in daily activities; it has become the most widely used clinical outcome measure for stroke clinical trials \cite{mRS1,mRS2}. In the state of the art, this problem of classification of patients according to the mRS was treated by classical ML approaches (Random Forest) by taking clinical variables \cite{van2018predicting} and radiological parameters (presence of leukoaraiosis, old infarctions, hyperdense vessel sign, and hemorrhagic transformation) \cite{ramos2020predicting}.

In this paper, we propose an approach which, in addition to the clinical metadata, takes the MRI images as input. This is achieved by proposing a spatio-temporal encoding which has already proved its efficiency in previous works \cite{giacalone2017,giacalone2018local}. However, unlike these previous works, this encoding is here specifically designed for deep learning architectures and is for the best of our knowledge presented for the first time in the context of images for stroke.
This spatio-temporal encoding is based on a convolutional neuronal network - long short-term memory (CNN-LSTM) architecture that involves using CNNs layers for spatial feature extraction on input MR images combined with LSTMs to support temporal sequence prediction. CNN-LSTM architecture was originally developed for image or video description \cite{venugopalan2016improving} and has very recently be proposed for problems in medical imaging for the classification of cancer types \cite{jang2018prediction,marentakis2021lung}.\\
In the context of our study, the CNN-LSTM architecture is applied independently to each MRI modality available at admission which are diffusion and perfusion MRI, more specifically we encode 5 entries which are raw diffusion MRI, the apparent diffusion coefficient (ADC) and the parametric maps associated to perfusion MRI that are time to maximum (Tmax), cerebral blood flow (CBF) and cerebral blood volume (CBV). 
In order to take advantages of ensemble learning (bias and variance reduction) \cite{hatami2012}, we then propose to fuse together the different classifiers constructed for each MR input by an ensemble voting approach. This fusion approach has the originality of being weighted using the clinical variables (not used as input in the CNN-LSTM architecture) to improve the classifiers' performance. This is done by giving higher weights (reward) to the modules that are in accordance with the clinical meta-data, and lower weights (punish) to those which are in disagreement. We exhaustively evaluated our framework based on a cohort of 119 patients with large intracranial artery occlusion treated by thrombectomy. Results show an accurate mRS prediction of 74\% for accuracy and 0.77 for AUC. We also compared our solution against different baselines. Our solution improves performance and stability by proposing an encoding adapted to the spatio-temporal nature of the image data and fusing together image and clinical data with an ensemble approach at the very end of the machine learning pipeline.\\

\section{METHODOLOGY}
\label{sec_methodology}


\begin{figure}
\centering
\centerline{\includegraphics[width=8.8cm]{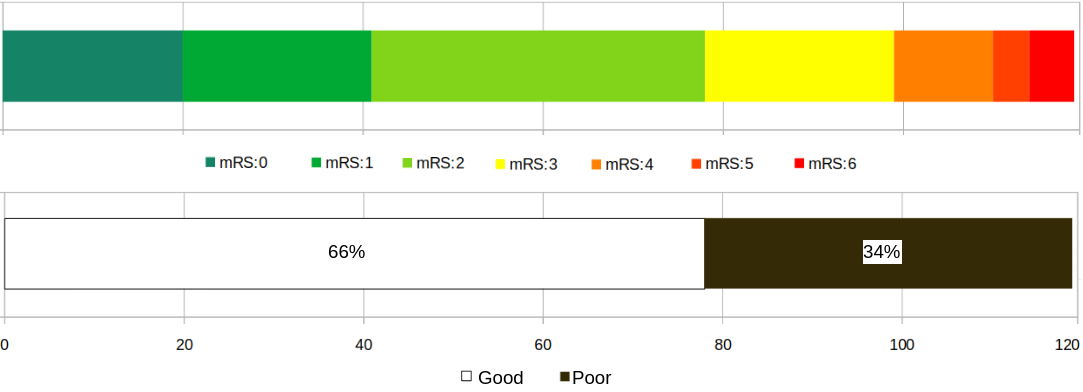}}
\caption{Top) Distribution of the 3-month mRS scores, bottom) associated binarization into good \{0,1,2\} and poor outcome \{3,4,5,6\}. 66\% vs. 34\% good and poor outcomes respectively within our cohort. }
\label{fig_mrsdata}
\end{figure}

\subsection{Data source and preprocessing}
Patients were included from the HIBISCUS-STROKE cohort \cite{Debs21}, which is an ongoing monocentric observational cohort enrolling patients with an ischemic stroke due to a proximal intracranial artery occlusion treated by thrombectomy. 
In total 119 patients, both male and female of age $67.20 \pm 16.32$ (mean$\pm$std), were analyzed. Inclusion criteria were: (1) patients with an anterior circulation stroke related to a proximal intracranial occlusion; (2) diffusion and perfusion MRI as baseline imaging; (3) patients treated by thrombectomy with or without intravenous thrombolysis. 
All patients gave their informed consent and the imaging protocol was approved by the regional ethics committee. 
All patients underwent the following protocol (IRB number:00009118) at admission: diffusion-weighted-imaging (DWI), dynamic susceptibility-contrast perfusion imaging (DSC-PWI) and a clinical evaluation including age and the National Institutes of Health Stroke Scale (NIHSS), which ranges from 0 to 42 (increasing scores indicate more severe neurological deficits) \cite{NIHSS14}. 
Final clinical outcome was assessed at 3-month during a face-to-face follow-up visit using the mRS. 
The distribution of the final mRS scores and its associated binarization into poor and good outcomes are shown in Figure \ref{fig_mrsdata}. In this paper, we used the binarized mRS for classifying patients' outcome.


Parametric maps were extracted from the DSC-PWI by circular singular value decomposition of the tissue concentration curves (Olea Sphere, Olea Medical, La Ciotat, France): cerebral blood flow (CBF), cerebral blood volume (CBV) and time to maximum (Tmax). DSC-PWI parametric maps were coregistered within subjects to DWI using linear registration with Ants (Avants et al., 2011) and all MRI slices were of size $192 \times 192$. The skull from all patients was removed using FSL (Smith et al., 2001). Finally, images were normalized between 0 and 1 to ensure inter-patient standardization.

\subsection{Proposed model and baselines}

\begin{figure}[htb]
\begin{minipage}[b]{1.0\linewidth}
  \centering
  \centerline{\includegraphics[width=8.6cm]{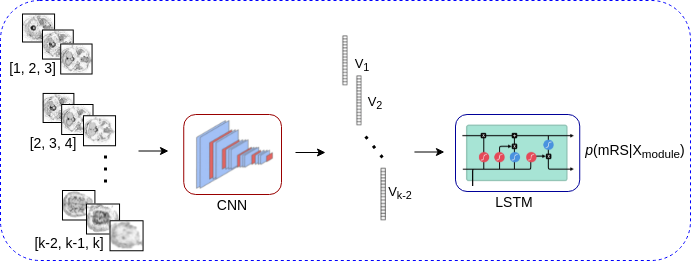}}
  \vspace{0.8cm}
\end{minipage}

\begin{minipage}[b]{1.0\linewidth}
  \centering
  \centerline{\includegraphics[width=8.6cm]{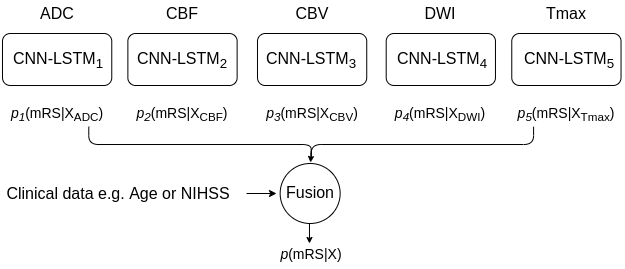}}
\end{minipage}
\caption{Top) The proposed CNN-LSTM block for a specific MR image encoding and bottom) the general block diagram of the proposed approach.}
\label{fig:mymodel}
\end{figure}


%

The general block diagram of the proposed multi-modal regression model for mRS prediction is shown in Figure \ref{fig:mymodel}. The image encoding part of the model (Figure \ref{fig:mymodel} bottom) consists of five modules ($N_\mathrm{modules}=5$) for ADC, CBF, CBV, DWI and Tmax. Each module receives sequences of images representing the entire parenchyma moving up the vascular tree from its lower part to its upper part. In order to extract the deep features and since the input for the pre-trained CNNs are RGB images, we used 3 consecutive slices. The resulting feature vectors were then used to train a LSTM \cite{LSTM97} for obtaining preliminary mRS predictions ($p(mRS|X_{module})$).

The final mRS prediction ($P_{mRS}$) is assessed using a weighted average of the preliminary probabilities, where the weights $\omega_{i}$ are calculated according to the Algorithm 1. The idea behind weighting with clinical data (i.e. age and/or NIHSS) is to reward the preliminary probabilities that are in agreement with the clinical data and penalize those in dissent. In other words, it uses the extra information provided by the clinical data to increase the confidence in the preliminary probabilities, and therefore in the final mRS prediction. The binary patient outcome is then predicted by setting a threshold on the final mRS prediction. 

Algorithm 1 includes a number of steps that we list hereafter. First, an optimal threshold is calculated for the image modules so as to maximize the AUCs (step 1). Then, the preliminary binary labels of the modules are calculated using the thresholds (step 2). Depending on the output of the modules (labels), a weight is assigned to them according to the clinical variable considered (step 3). Please note that the clinical variable we used here, i.e. age and NIHSS are positively correlated to the final output. In subjects with higher age or higher NIHSS score, the weights ($\omega_i$) reward the modules with "poor" outputs, and penalize the modules with "good" outputs. This way, the final mRS prediction is \emph{pushed} towards higher mRS values. These weights are then normalized to ensure that their sum is 1 on the scale of the 5 modules (step 4) and integrated with the preliminary probabilities in order to predict the final mRS (step 5). This prediction is then thresholded to produce the prediction of clinical outcome (step 6).

\begin{table} 
\centering
\begin{tabular}{p{8cm}}
\hline
\rowcolor{Gray}
\textbf{Algorithm 1}: Patient outcome prediction: fusion of the clinical meta-data and preliminary probabilities obtained from the image modules.\\ 
\hline
\vspace{-5pt}
\textbf{INPUT}: normalized clinical meta-data $c$, preliminary probabilities $p_{i}, i=1:N_{modules}$\\
\textbf{OUTPUT}: final mRS probability $P_{mRS}$ \\
\textbf{STEPS}:   \\
1. Find the global threshold $\tau$ for $p_{i}$ that maximize $AUC_{i}$\\
2. Obtain preliminary binary labels $l_{i} \in \{poor, good\}$\\
3. If $l_{i}$==\emph{poor}: \\
 \quad \quad \quad then: $\bar{\omega}_{i} = c$  \quad \quad \textbackslash\textbackslash penalize\\
 \quad  elseif $l_{i}$==\emph{good}:\\
 \quad \quad \quad $\bar{\omega}_{i} = 1-c$ \quad \quad \quad \textbackslash\textbackslash reward\\ 
4.  $\omega_{i} = \frac{\bar{\omega}_{i}}{\sum_{i=1}^{N_\mathrm{modules}} \bar{\omega}_{i}}$
\hspace{2mm}, where $\sum_{i=1}^{N_\mathrm{modules}}\omega_{i}=1$\\
5. $P_{mRS} =  \sum_{i=1}^{N_\mathrm{modules}} \omega_{i} p_{i}(mRS|X_{i})$\\
6. If $P_{mRS} \leq \tau^{*}$:\\
\quad \quad Final Patient Outcome $\leftarrow$ "good" \\
\quad else: \\
\quad \quad Final Patient Outcome $\leftarrow$ "poor" \\
\hline
\end{tabular} 
\label{algorithm1}
\end{table}

\begin{table*}[t]

\caption{Example of 5 patients that the proposed image-clinical data fusion algorithm corrects the prediction compared to the standard fusion of image modules only. Using age (top) and NIHSS (bottom) as clinical meta-data, respectively.}

\vspace{-0.5cm}

\small
\begin{center}
\begin{tabular}{llllll}
\begin{tabular}{c c c c c c c c c c c c c c c c}
\hline

Patient &  Age  & $p_{1}$ & $p_{2}$ & $p_{3}$  & $p_{4}$ & $p_{5}$ & $\omega_{1}$ & $\omega_{2}$ & $\omega_{3}$  & $\omega_{4}$ & $\omega_{5}$ & $L_{ens}$ & $L_{ens}$ & GS  & GS \\[2pt]

ID &   & ADC & CBF & CBV & DWI & Tmax &  &  &   & & &  w/o $\omega$ & age & outcome & mRS \\[2pt]

\hline\rule{0pt}{12pt} 033  & 45  & 0.45  & 0.42 &  0.46 & 0.27 & 0.64 & 0.16 & 0.16 & 0.16  & 0.34 & 0.16 & poor & \bf{good}  & good & 2\\

046 & 80  & 0.14  & 0.27 & 0.55 & 0.59 & 0.36 & 0.08 & 0.08 &  0.37 &  0.37 & 0.08 & good & \bf{poor} & poor & 3\\

162 & 55  & 0.50  & 0.48 & 0.46 & 0.54 & 0.16 & 0.19 & 0.19 & 0.19 & 0.19  & 0.23 & poor & \bf{good}  & good & 0\\

198 & 88  & 0.24  & 0.56 & 0.31 & 0.14 & 0.35 & 0.06 & 0.73 &  0.06 & 0.06  &0.06 & good & \bf{poor} & poor & 5\\

213 & 94  & 0.58  & 0.39 & 0.24 & 0.38 & 0.22 & 1 & 0 & 0  &  0 & 0 & good & \bf{poor} &  poor & 4\\
[2pt]
\hline
\end{tabular}
\end{tabular}
\end{center}

\vspace{0.1cm}

\begin{center}
\begin{tabular}{llllll}
\begin{tabular}{c c c c c c c c c c c c c c c c}
\hline

Patient &  \small{NIHSS} & $p_{1}$ & $p_{2}$ & $p_{3}$  & $p_{4}$ & $p_{5}$ & $\omega_{1}$ & $\omega_{2}$ & $\omega_{3}$  &$\omega_{4}$ & $\omega_{5}$ & $L_{ens}$  & $L_{ens}$ & GS & GS \\[2pt]

ID &   & ADC & CBF & CBV & DWI & $T_{max}$ &  &  &   &   & &  w/o $\omega$ & nihss  & outcome & mRS \\[2pt]
\hline\rule{0pt}{12pt} 023 &   8 & 0.68  & 0.75 & 0.74 & 0.07 & 0.24 & 0.13 & 0.13 & 0.13 & 0.3 & 0.3 & poor &  \bf{good} & good & 2\\

024 &  11 & 0.48  & 0.33 & 0.59 & 0.46 & 0.31 & 0.17 & 0.24 & 0.17 & 0.17 & 0.24 & poor &  \bf{good} & good & 2\\

027  &  26 & 0.31  & 0.46 & 0.17 & 0.17 & 0.20 & 0 & 1 & 0 & 0 & 0 & good &  \bf{poor} & poor & 3\\

033  &  5 & 0.45  & 0.42 &  0.46 & 0.27 & 0.64 & 0.12 & 0.12 & 0.12 & 0.51 & 0.12 & poor &  \bf{good} & good & 2\\

046 & 23 & 0.14  & 0.27 & 0.55 & 0.59 & 0.36 & 0.05 & 0.05 &  0.42 & 0.42 &0.05 &  good &  \bf{poor} & poor & 3\\
[2pt]
\hline
\end{tabular}
\end{tabular}
\end{center}

\label{table_0}
\end{table*}

Table \ref{table_0} reports two sub-tables which illustrate, for five patients each, the preliminary probabilities $p_{i}$, the weights calculated by Algorithm 1 $\omega_{i}$ and their impact on the final prediction $L_{ens}$. We also give to represent the decision without these weights $L_{ens}$ w/o and the gold standard GS. This table demonstrates how the final predicted outcome is corrected (shown in bold) by the proposed image-clinical data fusion algorithm, compared to the standard ensemble (fusion of image modules by averaging). For example (Table \ref{table_0}, bottom panel), patient 023 has an NIHSS score of 8 and "good" output (gold standard mRS$\leq$2) given by the clinicians. The preliminary probabilities of the five CNN-LSTM modules are [0.68, 0.75, 0.74, 0.07, 0.24] which result in the average ensemble probability of 0.49. Obtaining the best threshold ($\tau$) of 0.40 using the cross-validation, the final patient output for the ensemble is "poor". Given a low NIHSS for this patient, the proposed algorithm assigns the $\omega_{i}$ in a way that the final ensemble label $L_{ens}$ is corrected to "good". On contrary, patient 046 has an NIHSS of 23 and "poor" output (gold standard mRS$\geq$3). The high value of NIHSS help the algorithm to push the final score towards "poor", while the standard ensemble votes for "good" output. Please note that the proposed fusion algorithm uses only one clinical variable in an ensemble model, i.e. age or NIHSS. Age has a similar impact on the model's final output, as increasing age is associated with increasing likelihood for poor outcome (Table \ref{table_0}, upper panel).


There are three main hyperparameters to be tuned in the proposed model. First, the type of CNN chosen in the image encoding part: six popular CNN models VGG16, VGG19, Xception, ResNet50, MobileNet and DenseNet with pre-trained ImageNet weights were considered \cite{VGG16,ResNet}. Second, it concerns the hyperparameters related to the LSTM. Both normal and bidirectional LSTM is investigated, which we saw to significant difference among them. And finally, the threshold on the final mRS probability in order to obtain the binary labels. In order to determine the optimal hyperparameters, 5-fold cross-validation with patient-level separation is applied for maximizing AUC measure.




To compare our results, and because there is no published research on our dataset, we proposed three baseline models. First baseline is a Random Forest (RF) classifier inspired by \cite{ramos2020predicting}. 
The input for the RF are the following clinical data: NIHSS baseline, age, door-to-puncture-time and Fazekas scale. Second baseline is inspired by \cite{Debs21} where the input MR images are identical to our proposed model. It is an early fusion 3D-CNNs model (CNN), where the 5 MR images are concatenated and used as an unique input. 
In order to represent the architecture, we use $C_(size)$ and $F(size)$ where $C$ is a 3D convolution ($3\times3\times3$) followed by a ReLU activation function, a batch-normalization and 3D max-pooling layers and $F$ is fully-connected. Therefore, the CNN architecture is $C_(8)$-$C_(16)$-$C_(32)$-$C_(64)$-$C_(128)$-$C_(256)$-$F(100)$. 



\subsection{Evaluation}
Six different measures are used to evaluate the performance of the models: classification accuracy (recognition rate), F1 score, sensitivity, specificity, Mean-Absolute-Error (MAE) and the Area Under the Curve (AUC). 10 independent runs with random seeds are performed, and means and standard deviations of the measures are reported. For our specific binary problem with imbalanced classes (refer to the Figure \ref{fig_mrsdata}), some of the measures give more insight than the others. In mRS prediction, we believe AUC gives more precise evaluation of the overall binary models; and F1 score, sensitivity, and specificity also provide useful information as the false positive and false negative errors are critical for the considered medical application. MAE is generally used for evaluation of regression tasks and previously used to evaluate mRS prediction \cite{maier2017isles,winzeck2018isles}. 
In order to compare our model to the baselines we have also measured and reported the p-values of two-sided Wilcoxon signed rank tests. With p-value$\leq0.05$ we can reject the null hypothesis, and therefore the results predicted by two different models are significantly different with 95\% confidence. 

\section{EXPERIMENTAL RESULTS AND DISCUSSIONS}
\label{sec_experiments}


We carried out experiments in PyTorch. For each module, a LSTM regressor is trained to estimate the preliminary mRS scores from the feature sequences. The LSTM architecture consists of a sequence of input layers, two LSTM layers, a fully-connected and a sigmoid layer with the half-mean-squared-error loss. In order to obtain the best results, Adam optimizer with learning rate of $1e-4$ were tried with maximum number of epochs of $1000$ with early stopping and batch size of $32$. 
In our experiments, we explored different hidden nodes for LSTM $nh:\{50,100,500,1000\}$ and the best $nh:500$ using ResNet18 features was selected based on 5-fold cross-validation performance. 

Table \ref{table_1} reports the performance (mean$\pm$std over 10 runs) of the proposed ensemble compared to its individual modules. As shown, DWI is the best performing module (also performing well comparing to the ensemble) with AUC=0.71, MAE=0.38, F1 score=0.67, sensitivity=0.71, specificity=0.67, and accuracy=72\%. Regarding the interest of weighting vote with clinical data, both age and NIHSS have improved the ensemble performance, which proves the importance of integrating both imaging and clinical meta-data, and their complementary values to the prediction. It is also important to note that adding the NIHSS boosts the performance of our model more than with age. 


\begin{table*}[t]
\caption{Performance (mean and standard deviation over 10 runs) of the proposed ensemble compared to its individual modules.}

\vspace{-0.5cm}

\small
\begin{center}
\begin{tabular}{lllll}
\begin{tabular}{c c c c c c c c c}
\hline
 &  ADC & CBF & CBV & DWI & Tmax  & Ensemble & Ens. w-age & Ens. w-NIHSS\\[2pt]
\hline\rule{0pt}{12pt}
Acc.  & 0.69 $\pm$ 0.04 & 0.64  $\pm$ 0.03	& 0.59 $\pm $0.04  & 0.72 $\pm$ 0.03 &  0.65 $\pm$ 0.03 & 0.71 $\pm$ 0.04 & 0.72 $\pm$ 0.02 & \bf{0.74 $\pm$ 0.03}\\
Spec.  & 0.66 $\pm$0.03 & 0.63 $\pm$0.03 & 0.57$\pm$ 0.02  &	0.67 $\pm$ 0.04 & 0.59 $\pm$0.02  & 0.68 $\pm$ 0.03 &  0.68 $\pm$ 0.02 & \bf{0.69 $\pm$ 0.02}\\
Sens.  & 0.68 $\pm$0.05   & 0.62$\pm$ 0.03 & 0.57 $\pm$0.02  &0.70$\pm$ 0.04	& 0.60$\pm$ 0.04 & 0.68 $\pm$ 0.04  & 0.69 $\pm$ 0.02 & \bf{0.72 $\pm$ 0.05}\\
F1  &  0.66 $\pm$0.03 & 0.62$\pm$ 0.03 & 0.56$\pm$ 0.02 &	0.67 $\pm$0.04 & 0.59$\pm$ 0.02 & 0.68 $\pm$ 0.03 & 0.69 $\pm$ 0.02 & \bf{0.70 $\pm$ 0.02}\\
MAE & 0.39$\pm$ 0.01 & 0.40 $\pm$0.01 & 0.43 $\pm$0.01 & 0.38$\pm$0.01 & 0.41$\pm$0.00 & 0.40$\pm$0.00 & \bf{0.35$\pm$0.00} & \bf{0.35$\pm$0.00}\\
AUC & 0.69 $\pm$ 0.03   & 0.64 $\pm$ 0.03 & 0.56 $\pm$0.03  & 0.71 $\pm$ 0.04 & 0.58 $\pm$0.02  & 0.72 $\pm$ 0.03 & 0.74 $\pm$ 0.02 & \bf{0.77 $\pm$ 0.02}\\
[2pt]
\hline
\end{tabular}
\end{tabular}
\end{center}
\label{table_1}
\end{table*}

\begin{table}[t]
\caption{Performance (mean $\pm$ std over 10 runs) of the proposed ensemble compared to different baselines.}

\vspace{-0.5cm}

\small
\begin{center}
\begin{tabular}{lllllll}
\begin{tabular}{c c c c c }
\hline
 &  RF \cite{ramos2020predicting} & 3D-CNN & Ens.  & Ens wNIHSS\\[2pt]
\hline\rule{0pt}{11pt}
time &   \bf{$\sim$1 min}  & $\sim$90 min 	& $\sim$15 min  &	$\sim$15 min   \\
Acc. &   0.65$\pm$0.02   & 0.65$\pm$0.02 &  0.71$\pm$0.04 &  \bf{0.74$\pm$0.03}\\
Spec. &   0.64$\pm$0.02 &  0.64$\pm$0.02&  0.68$\pm$0.03 &	 \bf{0.69$\pm$0.02} \\
Sens. &   0.64$\pm$0.02   &  0.64$\pm$0.02	&  0.68$\pm$0.04 &	\bf{0.72$\pm$0.05} \\
F1  &  0.63$\pm$0.02    & 0.64$\pm$0.02	&  0.68$\pm$0.03 &	\bf{0.70$\pm$0.02}   \\
MAE  &  0.43$\pm$0.01  & 0.43$\pm$0.00 &  0.40$\pm$0.00 &	\bf{0.35$\pm$0.00}   \\
AUC &   0.65$\pm$0.02 & 0.66$\pm$0.02 	&  0.72$\pm$0.03 & \bf{0.77$\pm$0.02} \\
[2pt]
\hline
\end{tabular}
\end{tabular}
\end{center}
\label{table2}
\end{table}

Table \ref{table2} and \ref{table3} report the performance (mean$\pm$std over 10 runs) of the proposed ensemble (both with and without clinical meta-data) compared to the different baseline models. As shown, AUC, MAE, F1-score, sensitivity, specificity, accuracy of the proposed model is significantly superior than the baselines. Another advantage is
the training time, which is considerably reduced in the case of our model compared to the 3D-CNN model. 
Regarding the RF which is a shallow model, the saving in learning time and in the number of parameters comes at the cost of a significant drop in all the precision measurements. The low accuracy of the RF can have two explanations: 1) the RF inputs are hand-crafted features and cannot accept MRI images as a spatio-temporal data and 2) because of the limited number of parameters and RF depths, it tends to overfit and demonstrate a poor generalization ability (compared to the proposed model). 
It is also important to point out that the proposed ensemble with and without using clinical data outperform the baselines (see Table \ref{table2}, therefore the LSTM-CNN architecture is better suited for the MRI image coding than the RF and 3D-CNN.

From a statistical point-of-view, as shown in Table \ref{table3}, the results of the proposed models are statistically different from the baselines. It is also interesting to notice that the results from the baselines are similar, although they are different categories of algorithms and their input data is also different. 




\begin{table}[t]
\caption{\emph{P}-values of the proposed ensemble compared to the baselines.}

\vspace{-0.22cm}

\small
\begin{center}
\begin{tabular}{lllll}
\begin{tabular}{c c c c }
\hline
 &  RF vs. CNN & RF vs. Ens. & CNN vs. Ens. \\[2pt]
\hline\rule{0pt}{12pt}
Acc. &   0.826	&  \bf{0.002}	&  \bf{0.002}\\
Sens. &  0.625  &  \bf{0.002}  &  \bf{0.003} 	\\
Spec. & 0.492 &  \bf{0.002}	&  \bf{0.005}	\\
F1 &   0.675	& \bf{0.002}	& \bf{0.003}\\
MAE &    0.99	&  \bf{0.002} &  \bf{0.002} \\
AUC &    0.99	& \bf{0.002}	& \bf{0.002}		\\
[2pt]
\hline
\end{tabular}
\end{tabular}
\end{center}
\label{table3}
\end{table}


Comparing our results with the state-of-the-art, there are some important points to comment. First, our results offer a balanced prediction, meaning the prediction of good outcome is as good as the prediction of the poor outcome. While methods such as \cite{ramos2020predicting} focused only on the poor outcome and suffer a high false positive; in contrast, our model offers both high sensitivity and specificity.
Another interesting point about our results is the importance of age versus NIHSS in the outcome prediction. Our finding is in line with another independent research concluding that the NIHSS is likely to have a greater impact than age when it comes to mRS prediction \cite{amitrano2016}.
Lastly, as also shown in \cite{ramos2020predicting}, both age and NIHSS are important features for mRS prediction, and boost the prediction accuracy when combined to other predictor variables.\\
What is our explanation on NIHSS performing better than age in the
model? 
Although both increasing age and higher NIHSS scores are associated with worse clinical outcome \cite{andersen2011}, elderly patients can benefit from therapy (reperfusion), especially patients with milder baseline neurological severity \cite{Drouard2019}. In our dataset, the NIHSS score appears to add more prognostic information than age. It is likely related to the selection criteria of our patients (i.e. patients deemed eligible for thrombectomy), in whom the individual clinical severity has more prognostic importance than age. Elderly patients with significant pre-stroke comorbidities and disability are less likely to be treated by thrombectomy and thus included in the present study.

\section{CONCLUSIONS AND FUTURE WORK}
\label{sec_conclusions}

A CNN LSTM-based multimodal MR image fusion for predicting the final mRS is proposed. The proposed model offers the following advantages: (1) efficient encoding that fits the spatio-temporal nature of MRI data, (2) original fusion of MR images and clinical meta-data in a unified framework, 
(3) since the image deep features are extracted from the \emph{off-the-shell} CNNs (previously trained on the ImageNet \cite{ImageNet15}), the training part is only LSTMs which reduces the computational cost, while offering accuracy boost by comparison with the state-of-the art. 

One of the main limitations of the proposed fusion model is that it can only use one clinical variable in an ensemble model, e.g. either age or NIHSS. One future research is to adapt/extend the proposed weighting algorithm in order to combine multiple clinical variable with multiple image modules.

\section*{ACKNOWLEDGMENT}
This work was supported by the RHU MARVELOUS (ANR-16-RHUS-0009) of Universite Claude Bernard Lyon-1 (UCBL) and by the RHU BOOSTER (ANR-18-RHUS-0001), within the program ”Investissements d’Avenir“ operated by the French National Research Agency (ANR).


\bibliographystyle{IEEEbib}
\bibliography{refs.bib}

\end{document}